\documentclass[review]{elsarticle}

\usepackage{lineno,hyperref}
\usepackage{array}
\usepackage{amssymb}
\usepackage{enumerate}
\usepackage{makecell}
\usepackage{array}
\usepackage{multirow}
\usepackage{color}
\usepackage{diagbox}
\usepackage{enumerate}
\usepackage{graphicx}
\usepackage{subfigure}
\usepackage{CJK}
\usepackage{indentfirst}
\usepackage{amsmath}
\usepackage[justification=centering]{caption}
\usepackage{tabularx}
\usepackage{algorithm}
\usepackage{algorithmicx}
\usepackage{algpseudocode}
\usepackage{adjustbox}
\modulolinenumbers[5]
\newtheorem{definition}{Definition}[section]
\newtheorem{example}{Example}[section]

\newtheorem{property}{Property}

\newcommand{\tabincell}[2]{\begin{tabular}{@{}#1@{}}#2\end{tabular}}

\journal{Journal of \LaTeX\ Templates}

\begin{document}

\begin{frontmatter}

\title{Higher order information volume of mass function}

\author[mymainaddress]{Qianli Zhou}
\ead{zhouqianli@std.uestc.edu.cn}

\author[mymainaddress]{Yong Deng\corref{mycorrespondingauthor}}
\cortext[mycorrespondingauthor]{Corresponding author}
\ead{dengentropy@uestc.edu.cn}
\address[mymainaddress]{Institute of Fundamental and Frontier Science, University of Electronic Science and Technology of China, Chengdu 610054, China}

\begin{abstract}
For a certain moment, the information volume represented in a probability space can be accurately measured by Shannon entropy. But in real life, the results of things usually change over time, and the prediction of the information volume contained in the future is still an open question. Deng entropy proposed by Deng in recent years is widely applied on measuring the uncertainty, but its physical explanation is controversial. In this paper, we give Deng entropy a new explanation based on the fractal idea, and proposed its generalization called time fractal-based (TFB) entropy. The TFB entropy is recognized as predicting the uncertainty over a period of time by splitting times, and its maximum value, called higher order information volume of mass function (HOIVMF), can express more uncertain information than all of existing methods.
\end{abstract}


\end{frontmatter}

\section{Introduction}
\label{intro}
In order to deal with the uncertain events described by probability theory (PT), Shannon proposed information entropy called Shannon entropy in \cite{shannon2001mathematical}, which has satisfactory performance in describing the mutually exclusive information. For a probability distribution, Shannon entropy represents the information volume of it in the space and moment. For a random variable with the number of mutually exclusive events $n$, when it is uniformly distributed, it has the largest information volume $\log n$, which represents the maximum information volume that can be expressed by probability theory (PT) for $n$ mutually exclusive events in certain moment and space. But in real life, there are many events that cannot be linearly represented by probability distributions. On the space scale, Mandelbrot \cite{mandelbrot1967long} proposed the fractal theory; on the time scale,  Lorenz \cite{lorenz1955available} proposed the chaos theory. For processing incomplete mutually exclusive information, Zadeh \cite{zadeh1996fuzzy} proposed fuzzy set theory, and developed into Z number \cite{Jiang2019Znetwork,tian2020zslf}, D number \cite{liu2020anextended,IJISTUDNumbers}, intuitionistic fuzzy set \cite{Feng2020Enhancing}, Pythagoras fuzzy set \cite{Zhouqianli2020PFS}. They are widely used in medical diagnosis \cite{cao2019multi}, multi-criteria decision-making\cite{liao2020deng,fu2020multiple,FEI2020106355}, reliability analysis \cite{zhanglimao2019,liu2018classifier}.  Dempster and Shafer propose Dempster-Shafer evidence theory based on the multi-value probability mapping in \cite{dempster2008upper} and \cite{shafer1976mathematical}. Because the Dempster-Shafer theory (DST) can express more uncertain information than probability theory, it is applied on multi-source information fusion \cite{hurley2019nonlinear}\cite{wang2019improvement}, predicting interference effect \cite{Xiao2020CEQD},  pattern recognition \cite{zhu2017method} and classification decision \cite{liu2020evidence,Jiang2020MSML}, and further extended to complex evidence theory \cite{Xiao2019complexmassfunction,Xiao2020CED}.

How to measure the uncertainty of nonlinear system or incomplete mutually exclusive is an open issue \cite{cheong2019hybrid,Xiao2020GIQ,lai2020parrondo}. Zmeska \cite{zmeskal2013entropy} puts forward fractal entropy for fractal theory to describe the uncertainty of a nonlinear system, and  Lutz \cite{lutz2017bounding} uses fractal dimension to describe individual strings and sequences, which provides the possibility for the application of fractal ideas in information processing. In DST, many uncertainty measures have been proposed recently, the most widely applied of them including JS entropy \cite{jirouvsek2018new}, Deng entropy \cite{Deng2020ScienceChina} and SU uncertainty measurement \cite{wang2018uncertainty}. However, according to the total uncertainty measure requirements proposed by Kiler \cite{klir1990uncertainty} and Abell$\grave{a}$n \cite{abellan2008requirements}, Deng entropy \cite{Deng2020ScienceChina} produces undesirable results shown in \cite{abellan2017analyzing}\cite{moral2020critique}. SU uncertainty measurement \cite{wang2018uncertainty} has unreasonable results in \cite{zhou2020fractal} if we respectively discuss the discord and non-specificity parts proposed by Yager \cite{yager1983entropy}. In terms of reality, JS entropy \cite{jirouvsek2018new} has no physical model corresponding to Shannon entropy. Besides, the eXtropy\cite{dengeXtropy} and negation\cite{anjaria2020negation} also can be used in uncertainty measure. Based on above, we firstly combine fractal idea with belief entropy and proposed fractal-based (FB) entropy in \cite{zhou2020fractal}, It has better performance than all previous methods in measuring the uncertainty at a certain moment. But if we consider a period of time, FB entropy cannot be used to measure the uncertainty of basic probability assignment (BPA) in DST. So this paper proposes a generalized Deng entropy, called time fractal-based (TFB) entropy, which expresses the uncertainty over a period of time by splitting time to segments.

Similar to the information volume in probability theory (PT), in Dempster-Shafer theory (DST), Deng \cite{Deng2020InformationVolume} puts forward the information volume of mass function based on fractal idea and BPA of maximum Deng entropy, which is the first time to apply fractal idea on DST. But because of Deng's information volume splitting the BPA based on Deng entropy in a same proportion repeatedly, it cannot reach the maximum value in each order, and when the initial BPA or splitting proportion is changed, the maximum information volume also changes. The information volume corresponding to our proposed FB entropy \cite{zhou2020fractal} can represent the information volume of mass function at a certain moment. However, the uncertain information in reality changing with time, this paper, based on the TFB entropy,  proposes the higher order information volume of mass function (HOIVMF), which can express more uncertain information than Deng's method.

After the above introduction, the structure of paper is shown as follows:

\begin{description}
\item[$\bullet$]Some preliminary knowledge is introduced in the Section\ref{pre}, which can help readers understand the paper more easily.
\item[$\bullet$]In Section\ref{tfb1}. We give a new explanation of Deng entropy and introduced time fractal-based (TFB) entropy through a case. Then analyzing the properties of TFB entropy by some numerical examples.
\item[$\bullet$]Section\ref{hivmf} is the core of the paper, the maximum k-order TFB entropy is calculated and the mathematical reasoning is shown. The higher order information volume of mass function based on the maximum k-order TFB entropy is proposed.
\item[$\bullet$]Section\ref{c} summarizes the contributions of the paper, and puts forward the prospects for future research directions and contents.
\end{description}

\section{Preliminaries}
\label{pre}
This Section introduces the Dempster-Shafer theory (DST), Shannon entropy and its information volume, Deng entropy, FB entropy and other preliminary knowledge.

\subsection{Dempster-Shafer theory}
As a generalization of probability theory (PT), it can express more uncertain information than probability theory, whether in space or time. 
Hence, it has been well studied, including evidential reasoning \cite{zhou2018evidential,zhou2017evidential}, classification \cite{Liu2020Combination,fu2020comparison,liu2020evidencecombination}, industrial alarm system \cite{xu2018belief,xu2016optimal}, etc.

\begin{definition}[DST]
For a finite element set $\Theta=\{\theta_{1},\theta_{2},\dots,\theta{n}\}$, it is called a discernment framework to describe the state of evidence. Its power set $X=2^{\theta}=\{\varnothing,\{\theta_{1}\},\dots,\{\theta_{n}\},\{\theta_{1}\theta_{2}\},\dots\{\theta_{1}\dots\theta_{n}\}\}$ is composed with all subsets of $\Theta$, and each subset is called the focal element. Mass functions of them called basic probability assignment (BPA) are usually used to express the degree of support in the evidence, and the mass function should satisfy\cite{dempster2008upper}\cite{shafer1976mathematical}:
\begin{equation}
m(\varnothing)=0;~m(A)\ge0;~\sum_{A\in X}m(A)=1.
\end{equation}
\end{definition}
How to deal with the BPA of multi-element focal elements is the key to handle uncertainty of DST. Smets \cite{smets1994transferable} transforms BPA into probability distribution by average distributing, but this loses its non-specificity. Abell$\grave{a}$n et al. expresse BPA by using belief interval\cite{moral2020maximum}, but there are limitations in processing data. 
Some researchers focus on the BPA study of 
conflict coefficient \cite{Xiao2020Novel,Jiang2019IJIS},  
combination \cite{song2020selfadaptive,Fei2019Evidence}, etc.

\subsection{Information entropy}
Entropy is useful to measure uncertainty of information \cite{Xiao2020maximum}.
Especially, Shannon entropy has a wide range of applications in information theory, and it is now the most commonly used tool to express the uncertainty.
\begin{definition}[Shannon entropy]
For a n-dimensional random variable $P$, its probability distribution is $P=\{p_{1},p_{2},\dots, p_{n}\}$, and $\sum_{i=1}^{n}p_{i}=1$. The Shannon entropy of $P$ is defined as\cite{shannon2001mathematical}
\begin{equation}
H(P)=-\sum_{i=1}^{n}p_{i}\log p_{i}.
\end{equation}
When $\forall i = 1\rightarrow n$, $p_{i}=\frac{1}{n}$, the Shannon entropy reach the maximum $\log n$.
\end{definition}
The n-dimensional maximum Shannon entropy represents the maximum information volume of n-dimensional random variable expressed by the probability distribution. For the DST, we propose the fractal-based (FB) entropy based on the fractal idea.
\begin{definition}[FB entropy]
For a n-dimensional discernment framework $\Theta$, its power set is $X=2^{\Theta}$. The fractal-based entropy is defined as \cite{zhou2020fractal}:
\begin{equation}\label{fbentropy}
E_{FB}(\Theta)=-\sum_{F_{i}\in X} m_{F}(F_{i}) \log m_{F}(F_{i}),
\end{equation}
where $m_{F}(F_{i})$ is the representation of BPA in the dimension of probability distribution. It is defined as:
\begin{equation}
m_{F}(F_{i})=\frac{m(F_{i})}{2^{|F_{i}|}-1}+\sum_{F_{i} \subseteq G_{i} \cap |F_{i}| < |G_{i}|} \frac{m(G_{i})}{2^{|G_{i}|}-1}.
\end{equation}
If and only if $m(\Theta)=1$ or $m_{F}(F_{i})=\frac{1}{2^{|\Theta|}-1}$, the FB entropy reaches the maximum value $\log (2^{|\Theta|}-1)$.
\end{definition}

FB entropy can reasonably represent the information volume expressed by BPA at a certain moment. This method not only meets the requirements of total uncertainty measurement proposed by Kiler \cite{klir1990uncertainty} and Abell$\grave{a}$n \cite{abellan2008requirements}, but is superior to all existing methods as well. Deng proposes Deng entropy in \cite{Deng2020ScienceChina}, which is the first method to combine power set ideas with belief entropy.

\begin{definition}[Deng entropy]
For a n-dimensional discernment framework $\Theta$, its power set is $X=2^{\Theta}$. The Deng entropy is defined as \cite{Deng2020ScienceChina}:
\begin{equation}
E_{d}(\Theta)=-\sum_{F_{i}\in X}m(F_{i})\log \frac{m(F_{i})}{2^{|F_{i}|}-1}.
\end{equation}
In order to show more intuitively, Deng entropy can also be written as follows:
\begin{equation}\label{de}
E_{d}(\Theta)=-\sum_{F_{i}\in X}m^{\omega}_{d}(F_{i})\log m^{\omega}_{d}(F_{i}),
\end{equation}
where $m^{\omega}_{d}(F_{i})=\frac{m(\omega)}{2^{|\omega|}-1}(\omega\in X~and~F_{i}\subseteq\omega)$. If and only if $\forall F_{i}~m(F_{i})=\frac{2^{|F_{i}|}-1}{\sum_{G_{i}\in X}2^{|G_{i}|-1}}$ or $m^{\omega}_{d}(F_{i})=\frac{1}{\sum_{G_{i}\in X}2^{|G_{i}|}-1}$, the Deng entropy reaches the maximum value [] $\log (\sum_{F_{i}\in X}(2^{|F_{i}|}-1))$.
\end{definition}

\begin{figure}[htbp!]
\centering
\includegraphics[width=0.95\textwidth]{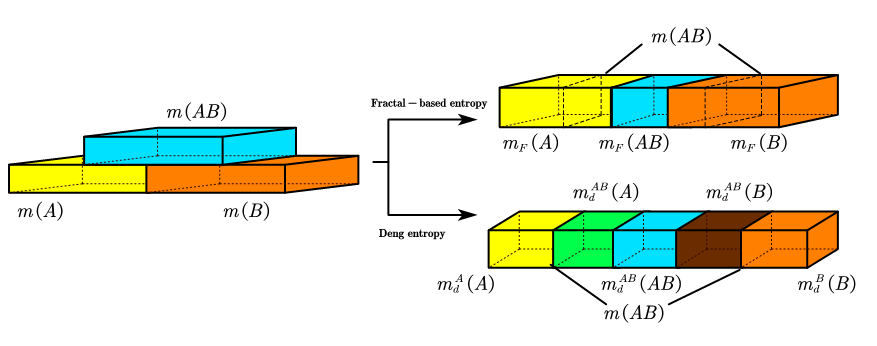}
\caption{FB entropy and Deng entropy's splitting of BPA}
\label{f2}
\end{figure}

Different from FB entropy, Deng entropy thinks that the subset after splitting of multi-element subsets cannot be directly added to the atomic subset, so Deng entropy can represent more information volume than FB entropy. Figure\ref{f2} shows the different operations of FB entropy and Deng entropy on BPA during splitting.

\subsection{Information volume of mass function}
According to the fractal idea and the maximum Deng entropy distribution, Deng proposes the information volume of mass function[].

\begin{figure}[htbp!]
\centering
\includegraphics[width=0.80\textwidth]{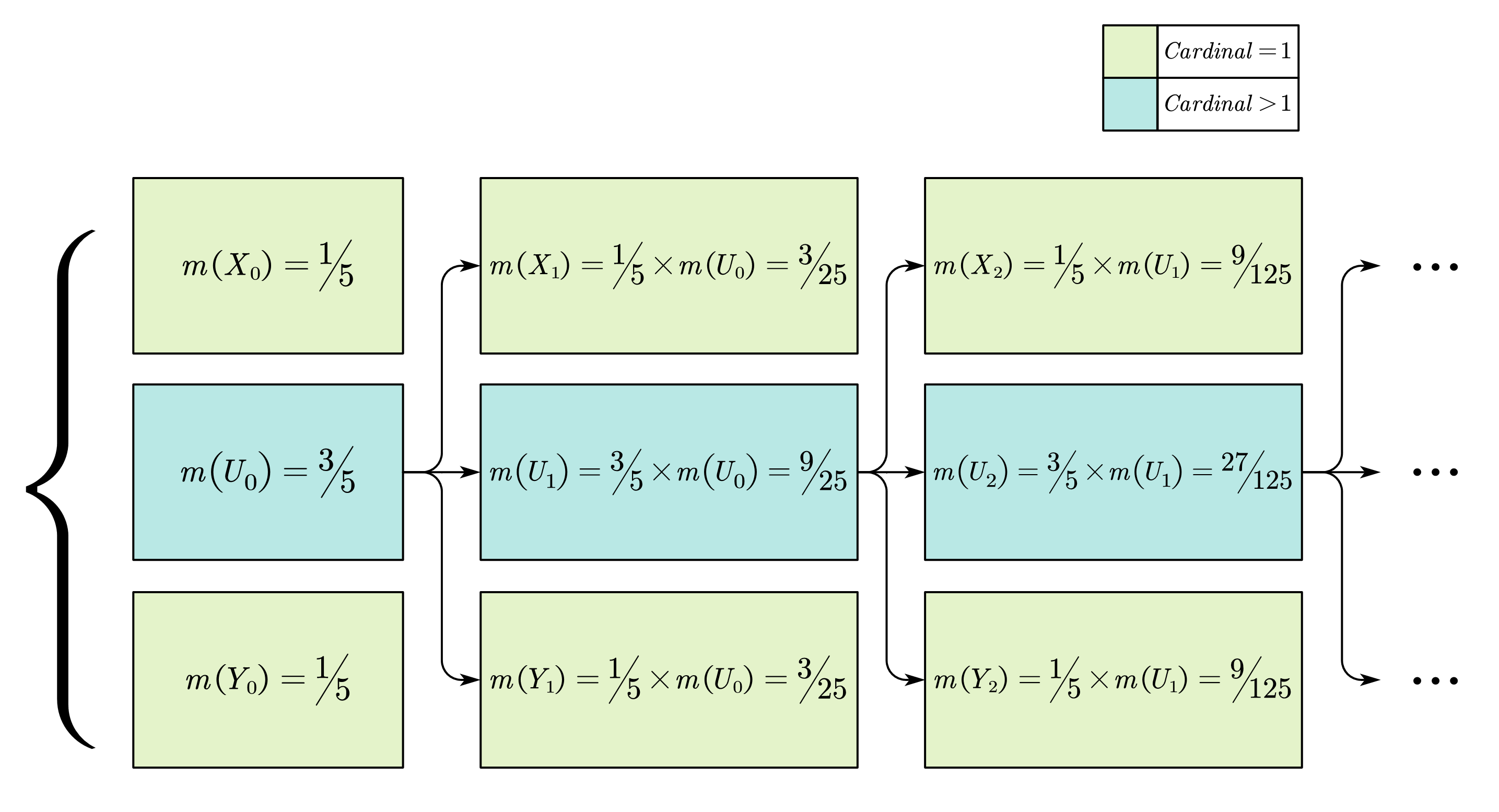}
\caption{Deng's information volume of mass function \cite{Deng2020InformationVolume}}
\label{f1}
\end{figure}

\begin{definition}[Information volume of mass function]\label{ivd}
For a n-dimensional discernment framework $\Theta$, its power set is $X=2^{\Theta}$. The information volume of mass function is defined by following steps[]:
\begin{description}
\item[$Step1$] Input the BPA $m(F_{i})$ of focal elements set $X$.
\item[$Step2$] Using the proportion of the maximum Deng entropy distribution is continuously split the mass function of the multi-element focal elements. Put the result of each split into Deng entropy formula until the increase of Deng entropy is less than $\epsilon$, where $\epsilon$ is error coefficient.
\item[$Step3$] Output the Deng entropy of the last iteration.
\end{description}
\end{definition}
In order to show the splitting process more intuitively, Figure \ref{f1} shows the corresponding splitting method of the 2-dimensional discernment framework $\Theta=\{X_{0},Y_{0}\}$.
The information volume of mass function proposed by Deng utilizes the idea of fractal for the first time, but it splitting the existing BPA, rather than the discernment framework, so the physical meaning of Deng's information volume is not clear yet.

\section{Time fractal-based belief entropy}
\label{tfb1}
In this section, a virus invasion case is used to give a new explanation to Deng entropy and generalized it to time fractal-based (TFB) entropy. And then the proposed entropy is compared with Deng entropy \cite{Deng2020ScienceChina}, FB entropy \cite{zhou2020fractal} and Shannon entropy \cite{shannon2001mathematical} to prove its necessary in uncertainty prediction.
\subsection{Speaking from the case of virus invasion}
\label{rb}
\begin{figure}[htbp!]
\centering
\includegraphics[width=0.98\textwidth]{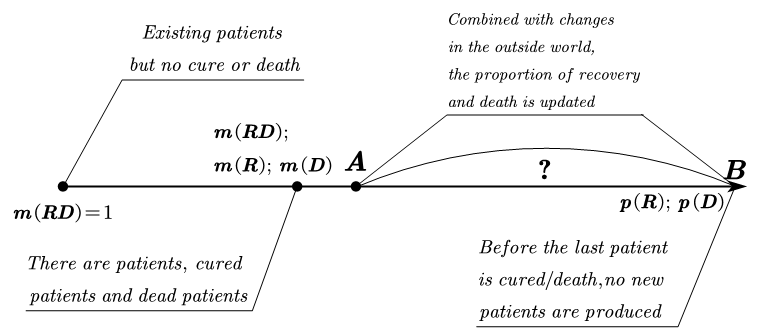}
\caption{The process of virus invasion}
\label{f3}
\end{figure}

Suppose a new human-to-human virus $C$ invades a city which has no population exchange with outside world. In the beginning, residents' treatment and prevention of the $C$ are unknown. With the time going by, residents’ medical ability and prevention methods for the $C$ have continued to improve. Until the end of the last patient's course, no new patients are generated, which means that the virus invasion is over. The entire invasion process is shown in the Figure\ref{f3}.
For the discernment framework $C=\{R,D\}$, $m(R)$ represents the cure rate, $m(D)$ represents the death rate, and $m(RD)$ represents the proportion of patients in the course of the disease. According to the Figure\ref{f3}, we know nothing about $C$ when no patient is cured or died, so $m(RD)=1$. The information volume at this moment can be expressed by FB entropy \cite{zhou2020fractal}, so the information volume at this moment is $\log(2^{2}-1)=1.5850$. When residents fight with the virus for a period of time, $m(R)$ and $m(D)$ at a certain moment are the results of the previous struggling with $C$. When the virus invasion is over (point $B$ and $m(RD)=0$),  $m(R)$ and $m(D)$ degenerate into a probability distribution, which is the result of the entire process of struggling with $C$. When we are at point $A$, how to predict the information volume for a period of time in the future (ex. segment $AB$), the existing method does not discuss this issue.

\subsection{The new explanation of Deng entropy}
\begin{figure}[htbp!]
\centering
\includegraphics[width=0.88\textwidth]{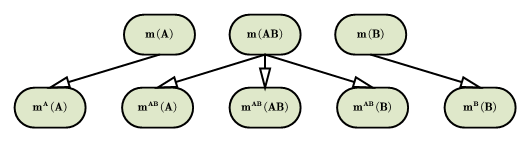}
\caption{The splitting method of Deng entropy \cite{Deng2020ScienceChina}}
\label{f31}
\end{figure}

Though Deng entropy proposed by Deng \cite{Deng2020ScienceChina} has satisfactory performance in many fields of uncertainty measure, its physical meaning and maximum value is controversial. Abell$\grave{a}$n et al. indicates that Deng entropy is not satisfied the monotonicity so that leading to undesirable results in some cases. In fact, it is unreasonable for Deng entropy to be used to measure BPA at a certain moment. According to the Equation\ref{de}, Deng entropy also can be calculated by the form of Shannon entropy. $m^{\omega}(\theta_{i})$ is the projection of BPA to the starting point in the evolution process, and the start point is original BPA. It is also can be seen as the BPA is split into its power set, and the mass functions of same focal elements after splitting is non-additive, because they are from different moments. For 2-dimensional discernment framework $\{A,B\}$, its splitting method is shown in Figure \ref{f31}, and Deng entropy of it is substituting the mass functions of second row to the Shannon entropy. And the BPA of maximum Deng entropy is composed with the uniform distribution of the second row's mass functions. Unlike other belief entropies, Deng entropy also covers a period of time while splitting the space, so the physical meaning of Deng entropy is predicting the information volume of segment $AB$ on the point $A$ of case in Section\ref{rb}.

Based on above, we can learn Deng entropy from a new perspective, and its mathematical explanation and physical meaning can reply the previous critiques of it.

\subsection{Time fractal-based entropy}
Deng entropy is one-step splitting of the multi-element focal elements, and it treats a period of time as a whole. But in reality, the time also can be splitting, and we can splitting time to many segments by continuously splitting the focal elements. According to this, k-order time fractal-based(TFB) entropy is calculated by continuously splitting the multi-element focal elements of their power set, and then uniformly distributing the original BPA into the mass functions of target $k$ order. Finally, substitute the mass functions of split focal elements into the Shannon entropy equation \cite{shannon2001mathematical} to obtain the result.

\begin{definition}[TFB entropy]\label{tfb}
For a n-dimensional discernment framework $\Theta=\{\theta_{1},\theta_{2},\dots,\theta_{n}\}$, its BPA set $M_{0}=\{m(\theta_{1}),m(\theta_{2}),\dots , m(\theta_{1}\dots\theta_{n})\}$. The k-order time fractal-based (TFB) entropy is defined as
\begin{equation}\label{tfbe}
E_{TFB}^{k}(M_{0})=-\sum_{F_{i}\in M_{0}}m(F_{i})\log \frac{m(F_{i})}{((k+1)^{|F_{i}|}-k^{|F_{i}|})},
\end{equation}
where order coefficient $k\in N^{+}$ and 1-order ATFB entropy is the Deng entropy[].
\begin{equation}\label{tfbe1}
E_{TFB}^{1}(M_{0})=-\sum_{F_{i}\in M_{0}}m(F_{i})\log \frac{m(F_{i})}{(2)^{|F_{i}|}-(1)^{|F_{i}|}}=E_{d}(M_{0}).
\end{equation}
\end{definition}

\begin{figure}[htbp!]
\centering
\includegraphics[width=0.98\textwidth]{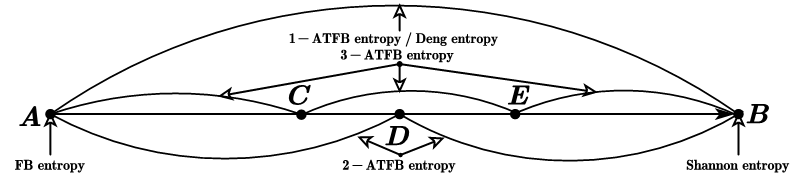}
\caption{1,2,3-order TFB entropy in Section \ref{rb}}
\label{f5}
\end{figure}

Observing the Definition \ref{tfb}, we can find that the TFB entropy is the generalization of the Deng entropy. So the physical meaning combining with Figure\ref{f3} of it can be explained as predicting the information volume of segments $AB$ on the point $A$. The order coefficient $k$ means splitting the $AB$ to $k$ segments to predict the uncertainty respectively. Figue\ref{f5} shows the relationship between $1,2,3$-order TFB entropy, Deng entropy, FB entropy and Shannon entropy in Section\ref{rb}.

\subsection{The numerical examples of ATFB entropy}
Some numerical examples are shown to illustrate the properties of ATFB entropy more intuitively.

\begin{example}\label{e1}
For 2-dimensional discernment framework $\Theta=\{A,B\}$, its BPA set $M_{0}=\{m(A)=\frac{1}{9},m(B)=\frac{1}{9},m(AB)=\frac{7}{9}\}$. The splitting process of $3$-order TFB entropy is shown in Figure\ref{e1f1}. For the mass function $m^{p}_{nq}(\omega)~(\omega \subseteq  \{A,B\})$ generated in the splitting process, in order to distinguish them reasonably, we make the following regulations:
\begin{description}
\item[$1.$] $p$ represents the $\omega$ from which focal element in the original BPA.
\item[$2.$] If $\omega$ is a single element, then $nq$ represents the nth-order multi-element focal element which splits itself into single element.
\item[$3.$] if $\omega$ consists of multiple elements, then $nq$ represents its last order focal element which splits itself into $\omega$.
\end{description}

\begin{figure}[htbp!]
\centering
\includegraphics[width=0.98\textwidth]{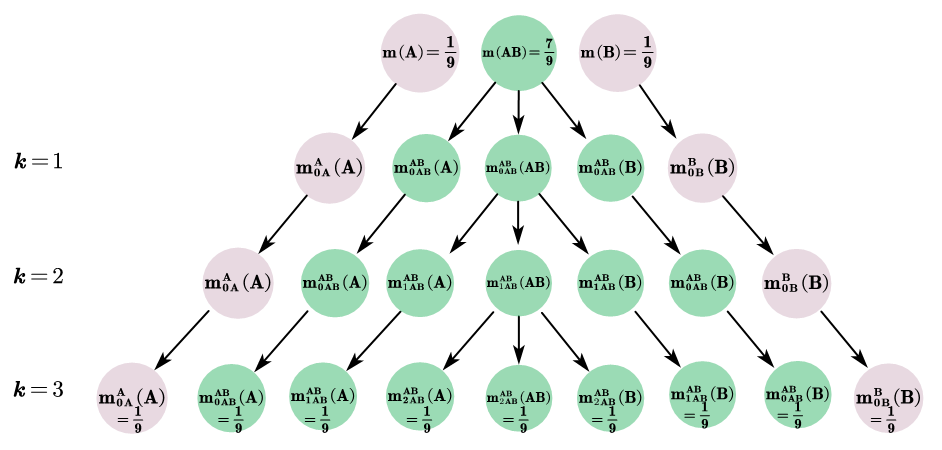}
\caption{The splitting method of TFB entropy in 2-dimensional discernment framework}
\label{e1f1}
\end{figure}

According to the mass function of last row, the $3$-order TFB entropy is $E_{TFB}^{k}(M_{0})=-9\times \frac{1}{9} \log (\frac{1}{9})=3.0294$.
\end{example}

Example\ref{e1} shows that for the k-order TFB entropy, the single element of k-order mass function is the same as BPA, and the multi-element focal elements' mass function is calculated by uniformly splitting the BPA. Though there are many splitting methods, the uniformly splitting can retain the maximum uncertainty to predict.

\begin{property}\label{p3}
With nothing know about the external environment, TFB entropy predicts the information volume for a period of time by retaining the maximum uncertainty of BPA.
\end{property}

\begin{example}\label{e2}
For a n-dimensional discernment framework $\Theta=\{\theta_{1},\theta_{2},\dots,\theta_{n}\}$, its BPA set $M_{0}=\{m(\theta_{1}),m(\theta_{2}),\dots , m(\theta_{1}\dots\theta_{n})\}$. Suppose $\sum_{i=1}^{n}m(\theta_{i})=1$, the k-order TFB entropy is
\begin{equation}
E_{TFB}^{k}(M_{0})=-\sum_{i=1}^{n}m(\theta_{i}) \log m(\theta_{i})=H(M_{0}),
\end{equation}
which is equal to the Shannon entropy of n-dimensional probability distribution.
\end{example}

Example \ref{e2} means that when BPA degenerates into probability distribution, TFB entropy also degenerate into Shannon entropy. This is because the matter at this time has a definite result, so TFB entropy can only represent the information volume at this moment. And the Property \ref{p1} can be described as follows.
\begin{property}\label{p1}
The probability distribution corresponding to the elements of discernment framework is the end point of TFB entropy prediction information volume.
\end{property}

\begin{example}\label{e3}
For a n-dimensional discernment framework $\Theta=\{\theta_{1},\theta_{2},\dots,\theta_{n}\}$, its BPA set $M_{0}=\{m(\theta_{1}),m(\theta_{2}),\dots , m(\theta_{1}\dots\theta_{n})\}$.
Suppose $m(\theta_{1}\dots\theta_{n})=1$, the k-order TFB entropy is
\begin{equation}\label{e3e}
E_{ATFB}^{k}(M_{0})=\log ((k+1)^{n}-k^{n}).
\end{equation}
\end{example}

By observing Equation\ref{e3e}, we can find that the increase of $k$ or $n$ can increase TFB entropy. For $n$, as the number of elements increases, the information volume increases intuitively.  For $k$, this is because that k-order TFB entropy is to predict the information volume of the period time by splitting time into $k$ segments to observe, which is like for observing a thing, comparing with hour-scale observation, minute-scale observation can find more information volume. So the Property \ref{p2} can be described as follows.
\begin{property}\label{p2}
For a n-dimensional discernment framework, the higher order TFB entropy means the higher the accuracy of predicting the information volume in the future.
\end{property}

\begin{example}\label{e4}
For a 2-dimensional discernment framework $\Theta=\{A,B\}$, its BPA set $M_{0}=\{m(A),m(B),m(AB)\}$. When $m(A)$ and $m(B)$ change in the interval $[0,1]$ respectively, the $1-9$-order TFB entropy is shown in the Figure\ref{e4f1}.

\begin{figure}[htbp]
\centering
\subfigure[k=1]{
\includegraphics[width=3.5cm]{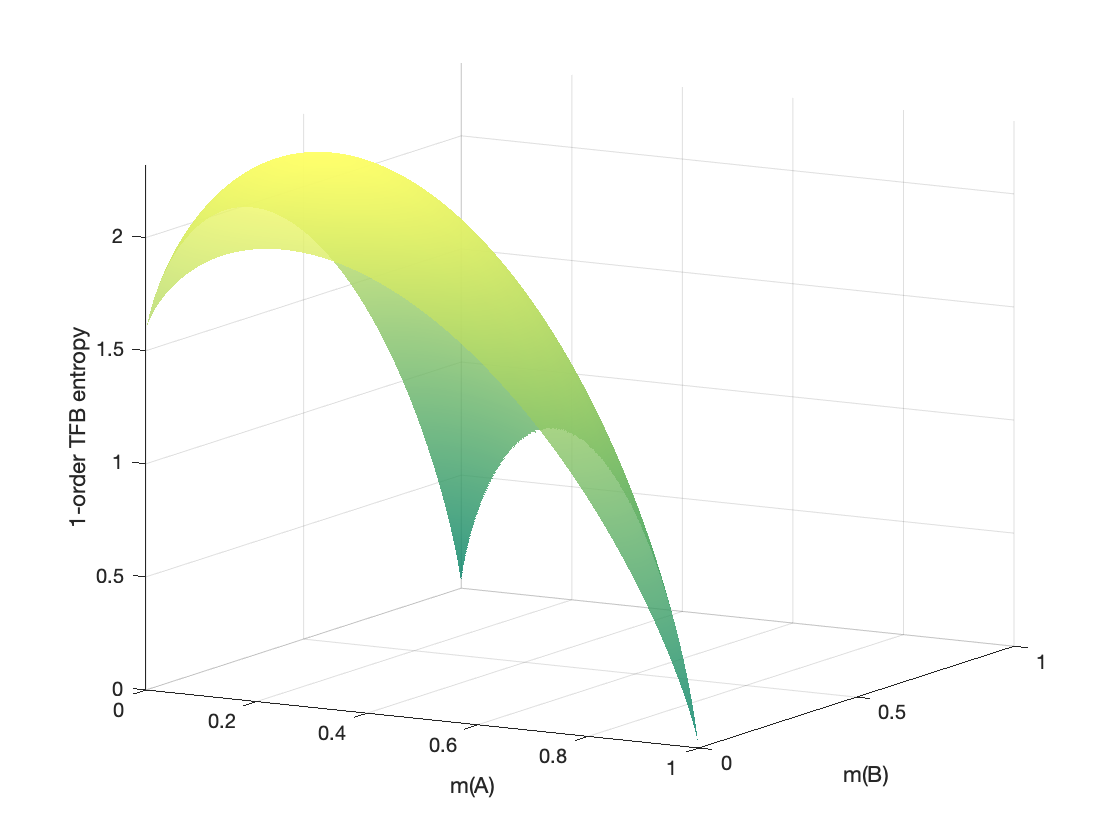}
}
\quad
\subfigure[k=2]{
\includegraphics[width=3.5cm]{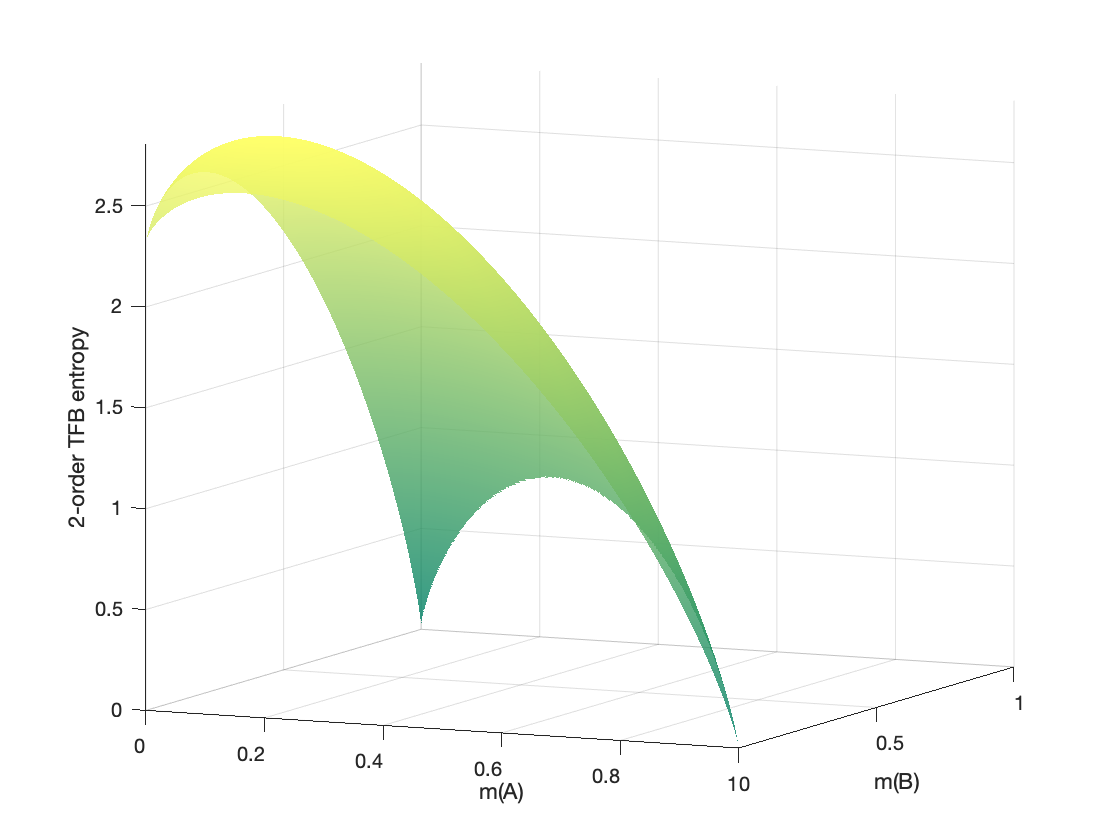}
}
\quad
\subfigure[k=3]{
\includegraphics[width=3.5cm]{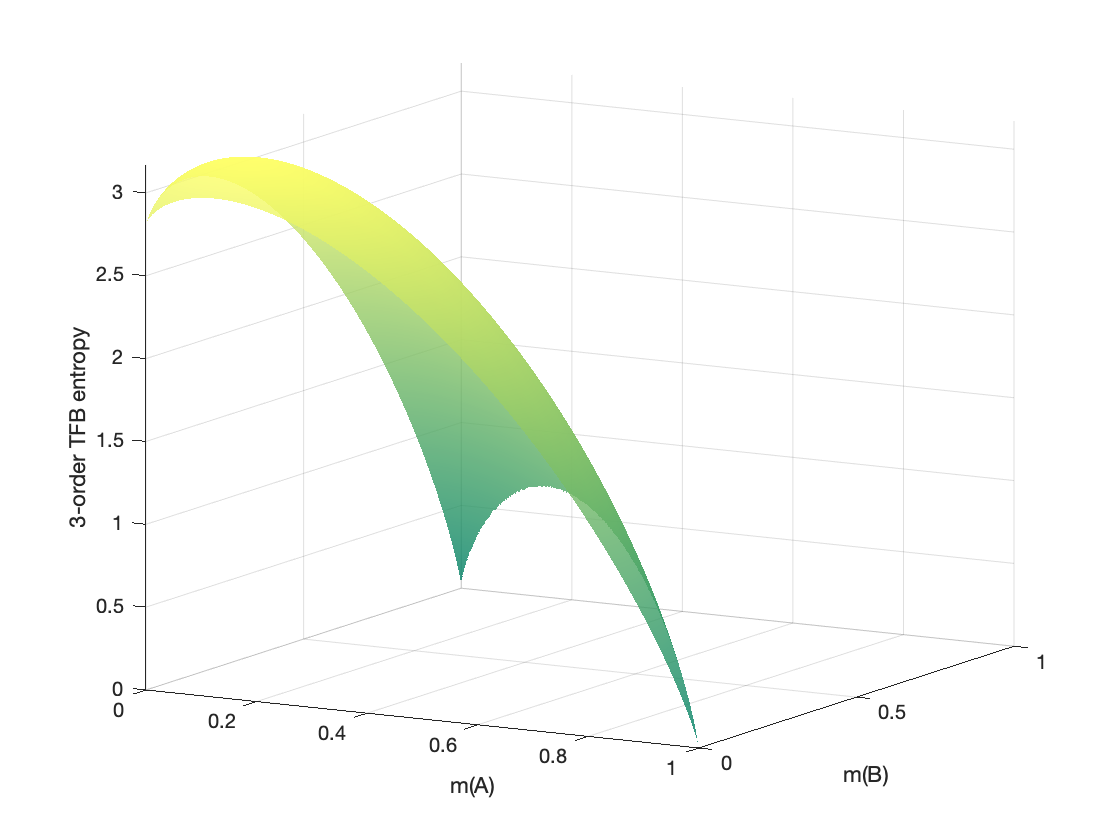}
}
\quad
\subfigure[k=4]{
\includegraphics[width=3.5cm]{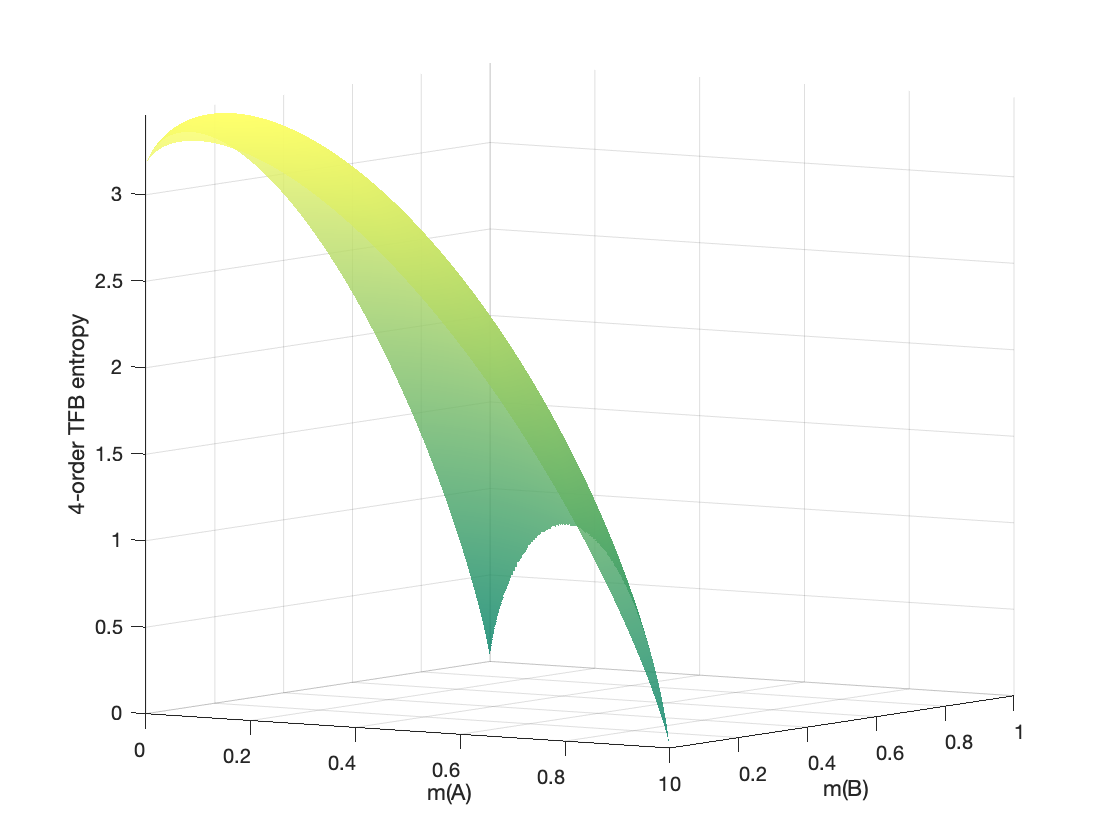}
}
\quad
\subfigure[k=5]{
\includegraphics[width=3.5cm]{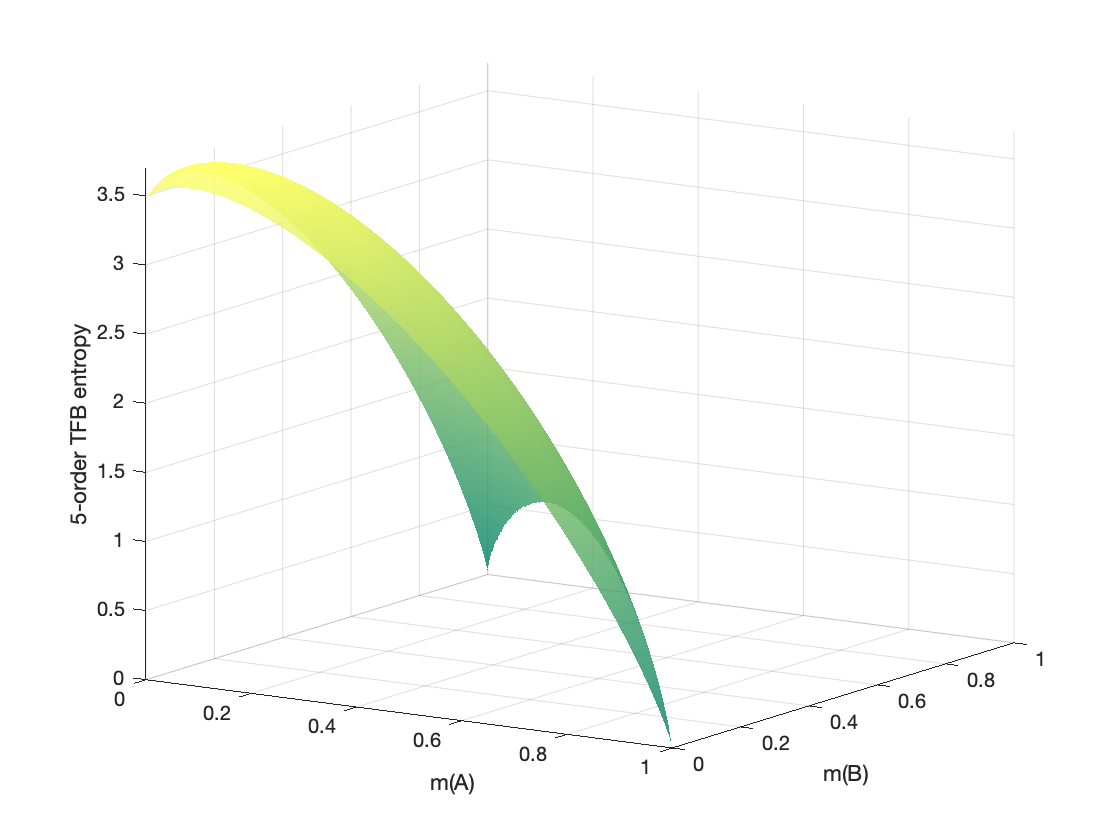}
}
\quad
\subfigure[k=6]{
\includegraphics[width=3.5cm]{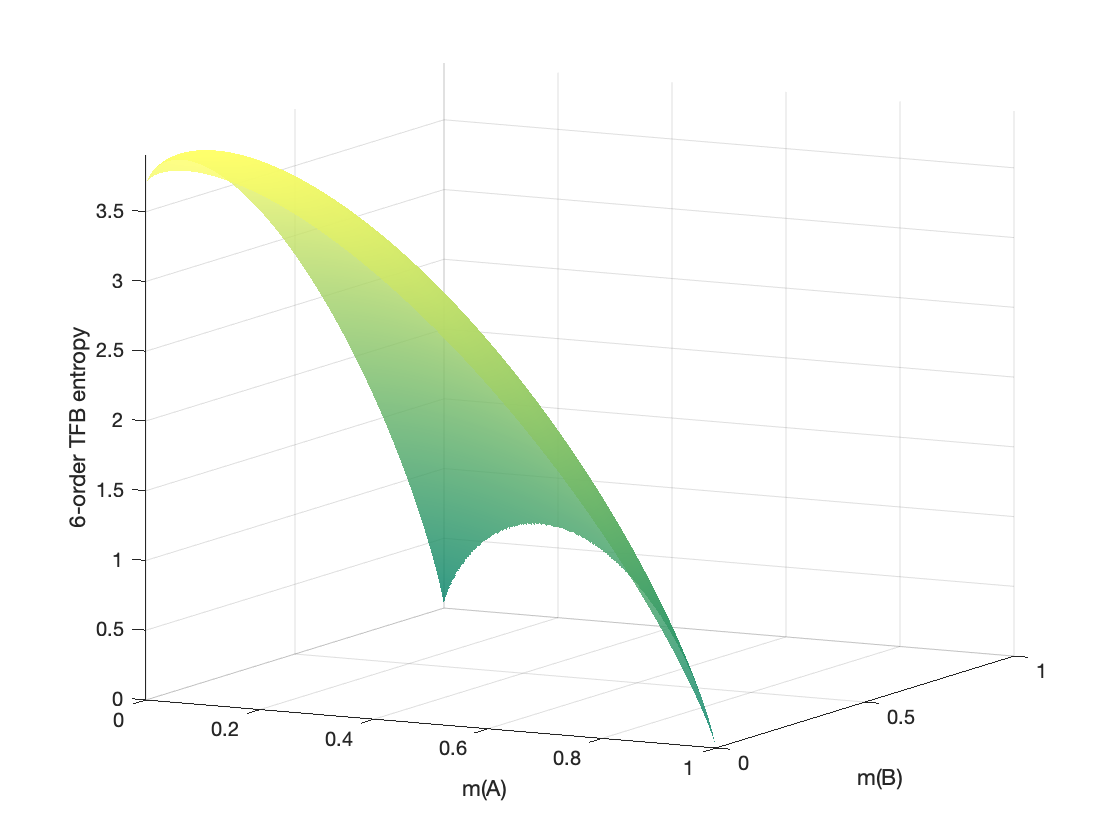}
}
\quad
\subfigure[k=7]{
\includegraphics[width=3.5cm]{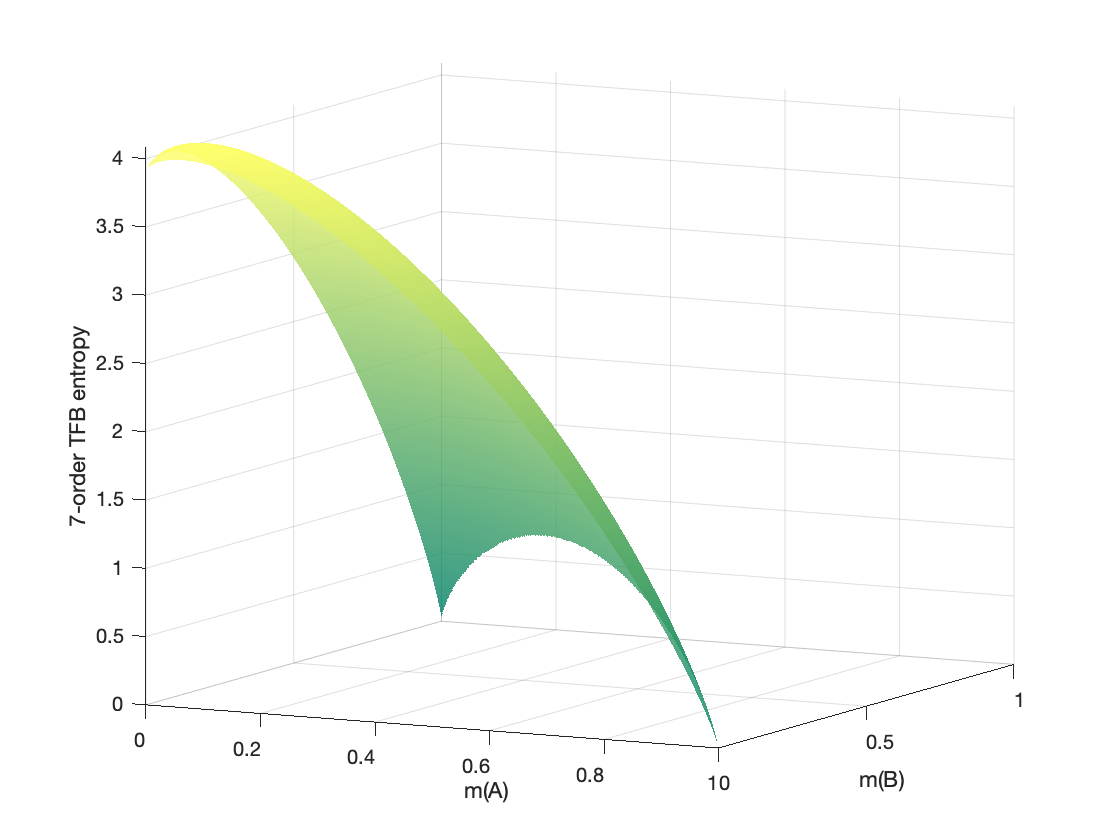}
}
\quad
\subfigure[k=8]{
\includegraphics[width=3.5cm]{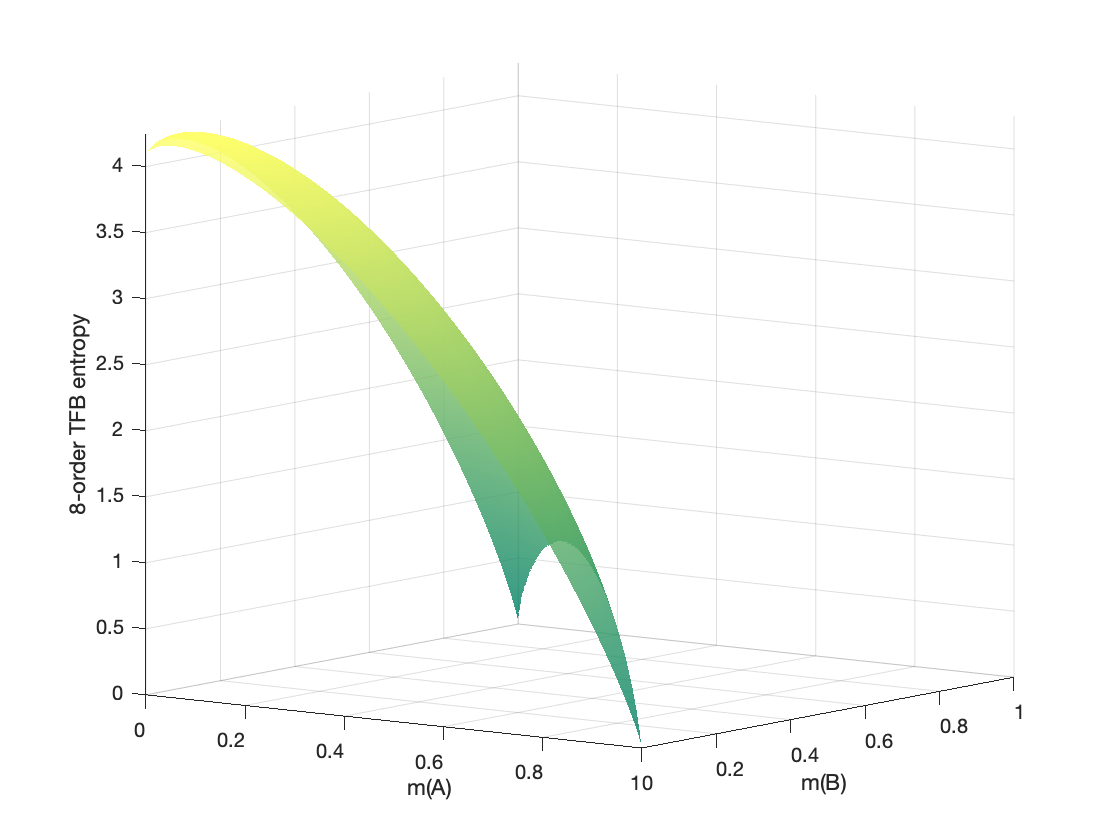}
}
\quad
\subfigure[k=9]{
\includegraphics[width=3.5cm]{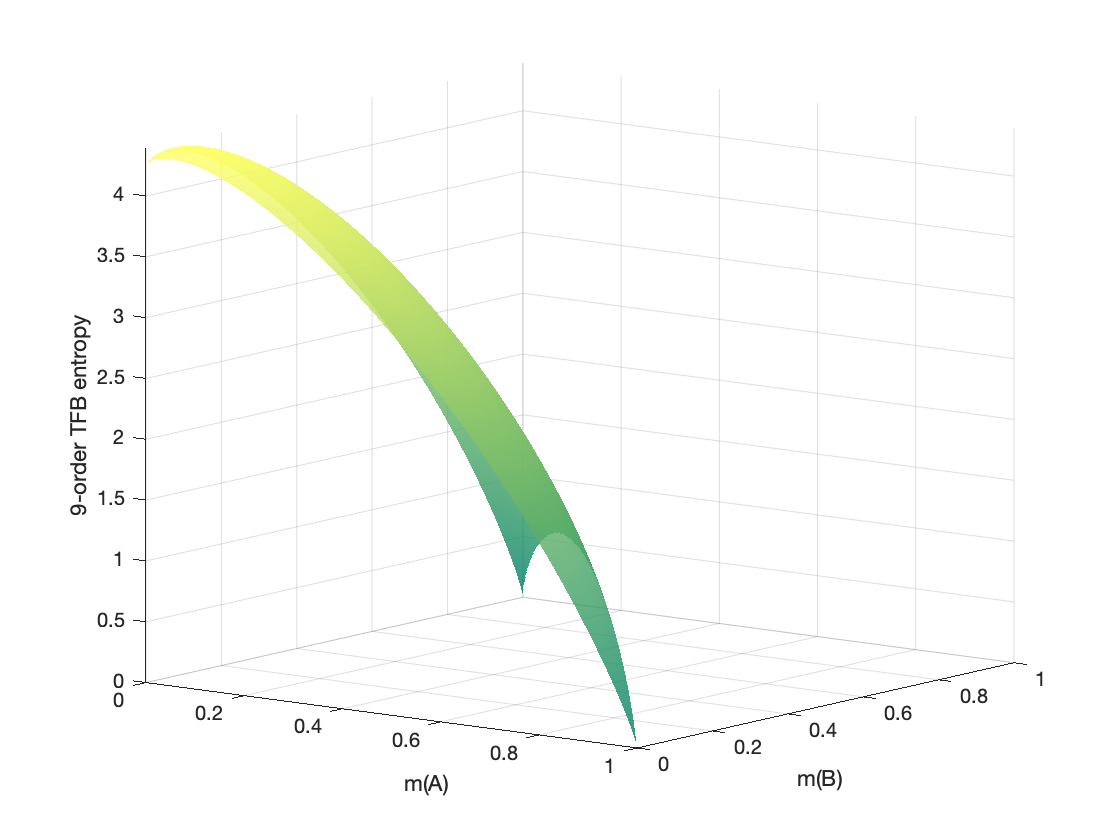}
}
\caption{$1-9$-order TFB entropy in Example\ref{e4}}\label{e4f1}
\end{figure}
\end{example}

Through the change trend of TFB entropies with $k$ in Figure\ref{e4f1}, it can be found that the value of the maximum k-order TFB entropy increases as the $k$ increases. And the BPA corresponding to the maximum k-order TFB entropy is also changing with the $k$, the trend is getting closer to $m(AB)=1$.

In this section, give Deng entropy a new explanation and generalized it to the TFB entropy, and utilize a virus invasion case .

\section{Higher dimensional information volume of mass function}
\label{hivmf}

Example\ref{e4} shows that in the 2-dimensional discernment framework, the BPA of maximum k-order FTB entropy is different with the $k$ increasing. In this section, we show the relationship between maximum k-order TFB entropy and its BPA and $k$. Expand the information volume of mass function proposed by Deng in \cite{Deng2020InformationVolume} based on Deng entropy.

\subsection{Higher dimensional information volume of mass function}
\label{hiv}
The maximum TFB entropy or the maximum information volume of $k$ order obtained when its mass functions reach the uniform distribution, and the value is $\log n$, where $n$ is the number of mass functions of the target order $k$.

\begin{example}\label{e5}
For a n-dimensional discernment framework $\Theta=\{\theta_{1},\theta_{2},\dots,\theta_{n}\}$, its BPA set $M_{0}=\{m(\theta_{1}),m(\theta_{2}),\dots , m(\theta_{1}\dots\theta_{n})\}$. For the different $n$ or $k$, their k-order maximum TFB entropy is shown in Table\ref{e5t1}.
\begin{table}[htbp!]\small
\caption{The maximum TFB entropy in Example \ref{e5}}
\label{e5t1}
\begin{center}
\begin{tabular}{c|cccc|c}
  \Xhline{1.4pt}
  Order&$\{\theta_{1}\theta_{2}\}$&$\{\theta_{1}\dots\theta_{3}\}$&$\{\theta_{1}\dots\theta_{4}\}$&$\{\theta_{1}\dots\theta_{5}\}$&$\{\theta_{1}\dots\theta_{n}\}$\\
  \hline
 $k=1$&$\log 5$ & $\log 19$ & $\log 65$ & $\log 211$ &$\log (\sum^{n}_{a=1}\dbinom{n}{a}(2^{a}-1^{a}))$\\
 $k=2$&$\log 7$ & $\log 37$ & $\log 175$ & $\log 781$ &$\log (\sum^{n}_{a=1}\dbinom{n}{a}(3^{a}-2^{a}))$\\
 $k=3$&$\log 9$ & $\log 61$ & $\log 369$ & $\log 2101$ &$\log (\sum^{n}_{a=1}\dbinom{n}{a}(4^{a}-3^{a}))$\\
 $k=4$&$\log 11$ & $\log 91$ & $\log 671$ & $\log 4651$ &$\log (\sum^{n}_{a=1}\dbinom{n}{a}(5^{a}-4^{a}))$\\
  \Xhline{1.4pt}
\end{tabular}
\end{center}
\end{table}

According to the right column, we can substitute the $k$ to the maximum TFB entropy $E_{TFB}^{k}(M_{n})$ to get

\begin{equation}
\begin{aligned}
E_{TFB}^{k}(M_{n})&=\log(\sum^{n}_{a=1}\dbinom{n}{a}((k+1)^{a}-k^{a}))\\
&=\log((\sum^{n}_{a=0}\dbinom{n}{a}(k+1)^{a}-1)-(\sum^{n}_{a=0}\dbinom{n}{a}(k)^{a}-1))\\
&=\log ((k+2)^n-(k+1)^n).
\end{aligned}
\end{equation}
\end{example}

From Example\ref{e5} we can find that the information volume of mass function being larger is because that the higher order mass functions have more dimensions than original BPA. So with the increasing of splitting order, we can define the higher order information volume of mass function (HOIVMF).

\begin{definition}[HOIVMF]
For a n-dimensional discernment framework $\Theta=\{\theta_{1},\theta_{2},\dots,\theta_{n}\}$, its k-order information volume of $(k+2)^n-(k+1)^n$-dimensional mass function is defined as follows,
\begin{equation}
E^{k}_{n}=\log ((k+2)^n-(k+1)^n).
\end{equation}
When the k-order FTB entropy equals to the k-order information volume, the original BPA's $m(\Theta)$ is
\begin{equation}\label{mme}
m(\Theta)=\frac{(k+1)^n-(k)^n}{(k+2)^n-(k+1)^n}.
\end{equation}
\end{definition}

According to Equation\ref{mme}, we can find that when $n$ is a constant, with the $k$ increasing, the $m(\Theta)$ is closer to $1$, which is consist with the trend in Figure\ref{e4f1}. When $k$ is infinite, $m(\Theta)=1$. So when the time is split into countless segments, the original BPA corresponding to the maximum entropy is $m(\Theta)=1$, which is intuitive. Therefore, when the scale of prediction is split to infinitely small, the TFB entropy corresponding to $m(\Theta)=1$ is the largest BPA. When the observation of things has been infinitely detailed,  $m(\Theta)=1$ is the most uncertainty BPA, which is the same as the result FB entropy \cite{zhou2020fractal} in a certain moment. From this perspective, the result of this splitting method is that the whole (TFB entropy) and the part (FB entropy) are similar, which is the core idea of fractal theory.

\subsection{Comparison with Deng's information volume of mass function}

Definition \ref{ivd} shows the Deng's information volume of mass function, which is splitting the BPA continuously in a same proportion. and then substitute the k-order mass function to Deng entropy. Because the same proportion splitting cannot reach the uniform distribution in $k$ order, even if it uses Deng entropy to calculate the information volume, the maximum information volume of Deng still smaller than proposed HOIVMF. we use a numerical example to display their relationship intuitively.

\begin{example}\label{e6}
For a 2-dimensional discernment framework $\Theta=\{A,B\}$, its BPA set $M_{0}=\{m(A),m(B),m(AB)\}$. Suppose $m(A)=m(B)=0.2;~m(AB)=0.6$. The Deng's $1-14$-order information value and the HOIVMF are shown in Table\ref{e6t1}.

\begin{table}[htbp!]\small
\caption{The information volume in Example \ref{e6}}
\label{e6t1}
\begin{center}
\begin{tabular}{c|ccc|cc}
  \Xhline{1.4pt}
  Order&\tabincell{c}{Deng's \\method}&\tabincell{c}{HOIVMF}&Order&\tabincell{c}{Deng's \\method}&\tabincell{c}{HOIVMF}\\
  \hline
$k=1$&$2.3219$ & $2.3219$ & $k=8$ & $3.3964$&$4.2479$\\
$k=2$&$2.7641$ & $2.8074$ & $k=9$ & $3.4088$&$4.3923$\\
$k=3$&$3.0294$ & $3.1699$ & $k=10$ & $3.4162$&$4.5236$\\
$k=4$&$3.1886$ & $3.4594$ & $k=11$ & $3.4206$&$4.6439$\\
$k=5$&$3.2841$ & $3.7044$ & $k=12$ & $3.4234$&$4.7549$\\
$k=6$&$3.3414$ & $3.9069$ & $k=13$ & $3.4250$&$4.8580$\\
$k=7$&$3.3758$ & $4.0875$ & $k=14$ & $3.4259$&$4.9542$\\
  \Xhline{1.4pt}
\end{tabular}
\end{center}
\end{table}
\end{example}

From Example\ref{e6}, we can find that because Deng entropy is only the 1-order ATFB entropy, which equals to the 1-order information volume. In higher dimensional information volume, our splitting methods can make the predicted information volume larger than the splitting method proposed by Deng \cite{Deng2020InformationVolume}, so the HOIVMF is superior to Deng's method.

\section{Conclusion}
\label{c}

This paper first proposed the belief entropy of a period time, called time fractal-based (TFB) entropy, which is the generalization of Deng entropy and can predict the future information volume at a moment. After verification, giving a BPA at a certain moment, the physical meaning of TFB entropy is predicting the information volume between the BPA and the probability distribution. The higher order information volume of mass function (HOIVMF) in $k$ order is defined as the maximum k-order TFB entropy. After the mathematical demonstration, for the n-dimensional discernment framework, the k-order information volume of mass function of it is $\log ((k+2)^n-(k+1)^n)$. Finally, we compared the HOIVMF with Deng's previous information volume and proved that the HOIVMF is more intuitive to predict the information.

In summary, the main contributions of this paper is shown as follows:

\begin{description}
\item[$1.$] This paper firstly proposes the information volume can be predicted/measured over a period of time, and a measurement method called TFB entropy based on fractal idea is given.
\item[$2.$] This paper firstly gives Deng entropy \cite{Deng2020ScienceChina} the physical meaning, and proposes Deng entropy's generalization, called ATFB entropy, which satisfies the mathematical operation law and intuition in predicting information volume for a period of time.
\item[$3.$] A new higher order mass function's information volume is proposed, and by comparing with the previous information volume proposed by Deng \cite{Deng2020InformationVolume} the new information volume can represent more uncertain information.
\end{description}

In the future research work, we further quantify the information volume for a period of time, and apply the prediction information volume in the fields of pattern recognition and information decision-making. Because a larger information volume and a longer time scale can help to achieve more accurate results in decision-making. In addition, we prepare to apply the idea of splitting time and fractal idea to more uncertainty measurements, because the changes in a period of time can describe things more accurately than a certain moment.

\section*{Declaration of interests}
The authors declare that they have no known competing financial interests or personal relationships that could have appeared to influence the work reported in this paper.

\section*{Acknowledgment}

The work is partially supported by the National Natural Science Foundation of China (Grant No. 61973332).

\section*{References}

\bibliography{mybibfile}

\end{document}